\documentclass[english]{article}
\usepackage{times}
\usepackage{url}
\usepackage{bbm}
\usepackage[T1]{fontenc}
\usepackage[latin1]{inputenc}
\usepackage{geometry}
\usepackage{rotating}
\usepackage{color}
\usepackage{graphicx}
\usepackage{amsmath, amsthm, amssymb}
\usepackage{setspace}
\usepackage{lineno}
\usepackage{natbib} \bibpunct{(}{)}{;}{author-year}{}{,} 

\usepackage{color}
\definecolor{trustcolor}{rgb}{0,0,1}

\makeatletter

\usepackage{babel}
\makeatother
\begin{document}

\title{Non-stationary patterns of isolation-by-distance: inferring measures of local genetic differentiation with Bayesian kriging.}

\author{Nicolas Duforet-Frebourg$^{1}$,  Michael G.B. Blum$^{1}$}

\date{~ }

\maketitle
$^{1}$ Universit\'e Joseph Fourier, Centre National de la Recherche Scientifique, Laboratoire TIMC-IMAG, Grenoble, France.


\noindent Running Head:  Inference of local genetic differentiation.

Corresponding author:
Michael  Blum

Laboratoire TIMC-IMAG, Facult\'e de M\'edecine, 38706 La Tronche, France 

Phone +33 4 56 52 00 65

Fax +33 4 56 52 00 55

Email: michael.blum@imag.fr

\clearpage

\begin{abstract}
{\normalsize 

Patterns of isolation-by-distance arise when population differentiation increases with increasing geographic distances. Patterns of isolation-by-distance are usually caused by local spatial dispersal, which explains why differences of allele frequencies between populations accumulate with distance. However, spatial variations of demographic parameters such as migration rate or population density can generate non-stationary patterns of isolation-by-distance where the rate at which genetic differentiation accumulates varies across space. To characterize non-stationary patterns of isolation-by-distance, we infer local genetic differentiation based on Bayesian kriging. Local genetic differentiation for a sampled population is defined as the average genetic differentiation between the sampled population and fictive neighboring populations. To avoid defining populations in advance, the method can also be applied at the scale of individuals making it relevant for landscape genetics. Inference of local genetic differentiation relies on a matrix of pairwise similarity or dissimilarity between populations or individuals such as matrices of $F_{ST}$ between pairs of populations. Simulation studies show that maps of local genetic differentiation can reveal barriers to gene flow but also other patterns such as continuous variations of gene flow across habitat. The potential of the method is illustrated with 2 data sets: genome-wide SNP data for human Swedish populations and AFLP markers for alpine plant species.}
\end{abstract}
\clearpage

\section*{Introduction}

Characterizing patterns of genetic differentiation within a species is a recurring task in population genetics. \citet{wright43} introduced the model of isolation by distance (IBD) which assumes that  differences of allele frequencies between populations accumulate under the assumption of local spatial dispersal. Because of local dispersal, IBD models predict the {\it pattern of IBD} where population differentiation increases with increasing geographic distances \cite[]{slatkin93, rousset97}. This pattern is observed in many model and non-model organisms as well as in humans suggesting that local dispersal is a leading evolutionary force \cite[]{sharbeletal2000,ramachandranetal05,hardyetal06,hellbergetal09}.

However, the pattern of IBD can mask complex variations of demographic parameters resulting in differential increases of genetic differentiation in different regions of the habitat. Variations of demographic parameters can arise when population densities or migration rates vary across space \cite[]{slatkin85}. With the advent of landscape genetics \cite[]{maneletal03,manel13}, the spatial variation of demographic parameters is an important topic because spatial heterogeneity (or landscape characteristics) is now recognized to be a key factor to explain population differentiation and gene flow \cite[]{mcraebeier07}. Examples of spatial heterogeneity influencing  population differentiation include varying local subpopulation size \cite[]{serrouyaetal12} as well as fragmented landscapes in urban and agricultural area where there are `corridors' for gene flow \cite[]{arnaud03,munshi12}. Barriers to gene flow, which can be caused by anthropogenic or geographic factors, are also emblematic examples of spatial heterogeneity influencing population structure \cite[e.g.][]{castellaetal00,eppsetal05,rileyetal06,gauffreetal08,zalewskietal09}. Because the identification of barriers to gene flow has attracted considerable attention \cite[]{storferetal10}, there is a large variety of statistical methods to detect them \cite[]{barbujanietal89,bocquetbacro94,dupanloupetal02,mannietal04,cercueiletal07,cridaetal07,maneletal07,safneretal2011}.  Here, we propose a more general method that characterizes  {\it non-stationary} patterns of isolation-by-distance. A  non-stationary pattern of isolation-by-distance occurs when the rate at which differentiation between individuals or populations accumulates with distance depends on space. Non-stationary patterns of IBD arise for instance when there is a barrier to gene flow because genetic differentiation accumulates more rapidly with distance around the barrier but they can also occur on different situations such as continuous variations of gene flow across the species range.

\begin{figure}[ht!]
	\centering
		\includegraphics[width=14cm]{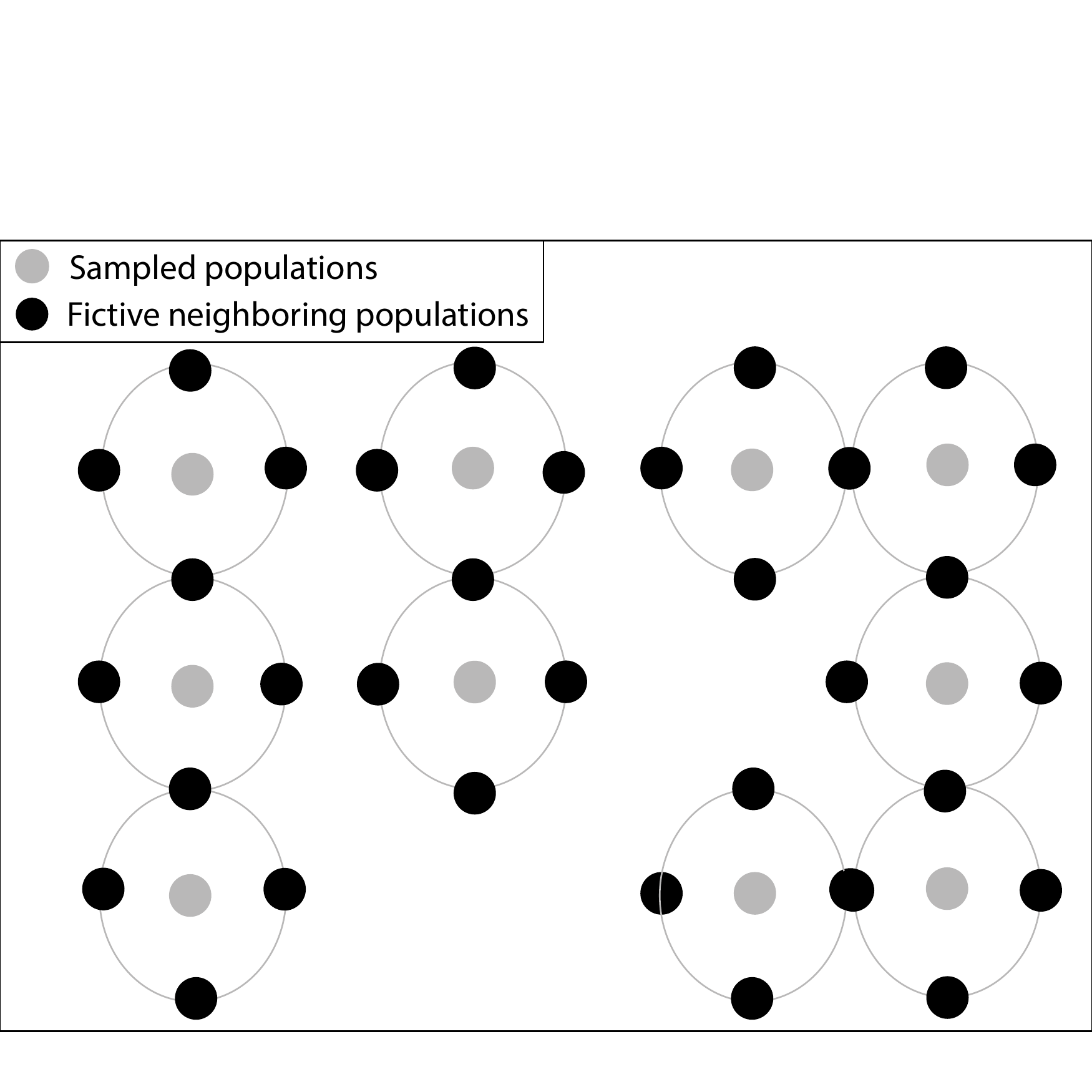}
	\caption{A 2-dimensional range with putative sampled locations (in grey) and neighboring locations (in black). Local differentiation at each sampled location corresponds to the average (over neighbors) pairwise measure of differentiation between the population or individual at the sampled location and its neighbors.}
 \label{fig:principle} 
\end{figure}

To characterize non-stationary patterns of IBD, our approach provides a measure of local differentiation at each location where genetic data are available. 
The principle of the method is to estimate for each sampled location $z_i$, $i=1,\dots,n,$ a local pairwise measure of population differentiation or of dissimilarity between the population located at the sampled location and fictive neighboring populations located at a fixed distance $d$ of $z_i$ (see Figure \ref{fig:principle}). Considering for instance $F_{ST}$ as a pairwise measure of genetic differentiation, the method provides estimates of $F_{ST}$ for pairs of populations separated by a distance $d$ and located in the vicinity of the sampling sites. The distance $d$ has to be set in advance and should be small compared to the dimension of the region under study. Fictive neighboring populations are introduced as a mean to provide measures of local genetic differentiation---$F_{ST}$ between populations separated by a distance $d$ here---that are comparable between sampling sites. Compared to common tests for isolation-by-distance \cite[]{hardyvekemans99}, the method is more informative because it quantifies how local genetic differentiation varies across space; the rate at which genetic differentiation increases with distance may vary across space and the proposed approach provides a quantitative assessment of this variation. To determine if variation of local differentiation is sufficiently large to reject stationary IBD, we additionally provide an hypothesis-testing procedure based on simulations. The method is not restricted to pairwise population measurements and can also accommodate individual pairwise measures. Working at the scale of individuals is a desirable feature since using individuals as the operational unit avoids potential bias in identifying populations in advance and offers the opportunity to conduct studies at a finer scale \cite[]{maneletal03,maneletal07}.  Using a detailed simulation study, we demonstrate that the method can correctly infer local variation of genetic differentiation and we present applications to human SNP data \cite[]{humphreysetal11} and AFLP markers from alpine plants \cite[]{gugerlietal08}.

\section*{Methods}

For the sake of the presentation, we assume that the data consist of allele frequencies in each population and that the method relies on the empirical correlation matrix between populations. In the RESULTS section, we show that the proposed approach is also appropriate with other pairwise matrices such as $F_{ST}$ matrices between populations or correlation matrices between individuals.

To assess local genetic differentiation around a given sampled site, we estimate the correlation of allele frequencies between the sampled population and  fictive populations located in the neighborhood of the sampling sites. Neighboring populations are located at a fixed and short distance from the sampled populations, and we measure the expected local correlation (averaged over neighbors) of allele frequencies between the sampled population and the neighboring fictive populations (Figure \ref{fig:principle}). Since we aim at providing local genetic differentiation values that should be larger in regions of abrupt genetic changes, we consider one minus the local correlation as a measure of local differentiation.

We estimate local correlation using a Gaussian process approach \cite[]{bishop06}, which is known as {\it kriging} in geostatistics \cite[]{cressie93}. Kriging refers to a set of interpolation methods where a variable of interest is estimated at unsampled locations based on values measured at the sampling sites. Interpolation relies on a weighted average of the values measured at the sampling sites and the weights depend on a parametric function $C$ which describes how the correlation or the covariance decreases with distance \cite[]{cressie93}. A direct application of kriging would consist of interpolating the allele frequencies at the neighboring sites based on the allele frequencies estimated at the sampled sites. However, the proposed approach is non-standard and requires methodological developments because we rather aim at estimating the correlation matrix between sampled and unsampled neighboring sites based on the correlation matrix between sampled sites. There is a vast literature of kriging procedures with non-stationary covariance when the function $C$ describing the decay of correlation with distance varies in space \cite[]{nottdunsmuir02,schmidtandohagan03,paciorekschervish06}. The covariance between sampled and unsampled sites is usually  estimated using a  parametric model \cite[]{paciorekschervish06} or at least using a given functional model for the covariance function \cite[]{schmidtandohagan03}. However, compared to geostatics where only one or a few variables are observed at the sampling sites, we are in a favorable situation in population genetics to estimate how the covariance or the correlation varies across space. Because each locus is a statistical replicate, there is enough information to estimate the correlation between the sampled sites using the empirical correlation matrix for instance. Estimating local correlation amounts at interpolating the correlation between sampled and neighboring sites from the correlation matrix between sampled sites. We explain below how we perform the interpolation step.


\subsection*{The kriging/Gaussian process approach}
In the following, we denote by $\bf{X}$ and $\bf{Y}$ the vectors of allele frequencies at sampled and unsampled sites. We assume independence between loci and the vectors $\bf{X}$ and $\bf{Y}$ contain allele frequencies for an arbitrary locus. The objective of the kriging approach is to interpolate the covariance (or correlation) matrix between $\bf{X}$ and $\bf{Y}$ based on the empirical covariance matrix between sampled sites. The covariance matrix between $\bf{X}$ and $\bf{Y}$ is denoted $\bf{E}[(\bf{Y}-\bf{m})(\bf{X}-\bf{m})^T]$ where $\bf{m}$ is a constant mean over the range. The main principle is to use weighted means of covariance values between sampled sites to estimated covariance between sampled and unsampled sites. As usual in kriging, the weights depend on a parametric function $C$ that gives the decay of correlation with distance. We explain below how we compute these weights.

The Gaussian process viewpoint of kriging is to model  the joint values of the variable at sampled and unsampled  sites as a multivariate Gaussian variable \cite[]{bishop06}
\begin{equation}
\label{eq:model}
(\bf{X},\bf{Y})\leadsto \mathcal{N}(\bf{m},\bf{\Psi}),
\end{equation}
where 
$$\bf{\Psi} = \left(
\begin{array}{cc}
\bf{\Psi_{xx}} & \bf{\Psi_{xy}} \\
\bf{\Psi_{xy}}^T & \bf{\Psi_{yy}} \\
\end{array}
\right),
$$
where $\bf{\Psi_{xx}}$ (resp. $\bf{\Psi_{yy}}$) denote the covariance matrix between the sampled sites (resp. unsampled sites) and $\bf{\Psi_{xy}}$ contain the covariances between the sampled and unsampled sites. The interpolation of the variable of unknown allele frequencies $Y$ is obtained using the conditional distribution of $\bf{Y}$ given $\bf{X}$, which can be written in the following regression form 
\begin{equation}
\label{eq:reg_krig}
\bf{Y}-\bf{m}= \boldsymbol \tau_{\bf \Psi} (\bf{X}-\bf{m}) +\boldsymbol \epsilon,
\end{equation}
where $\boldsymbol \tau_{\bf \Psi}=\bf{\Psi_{xy}}^T\bf{\Psi_{xx}}^{-1}$ and $\boldsymbol  \epsilon$ is a residual independent of $\bf{X}$ \cite[]{brownetal94}. A naive computation of local covariance would consist of simulating with equation (\ref{eq:reg_krig}) the vector $\bf{Y}$ containing the allele frequencies at the neighboring sites and then to evaluate numerically the empirical covariance between allele frequencies at sampled and at neighboring sites. Although it is a valid approach, we can actually derive what is the expected covariance between sampled and neighboring sites using equation (\ref{eq:reg_krig}) and we obtain
\begin{equation}
\bf{E}[(\bf{Y}-\bf{m})(\bf{X}-\bf{m})^T]=\boldsymbol \tau_{\bf \Psi} \bf{E}[(\bf{X}-\bf{m})(\bf{X}-\bf{m})^T]. 
\label{eq:covariance}
\end{equation}

In the computations, we replace $\bf{m}$ by the empirical mean  so that we estimate the covariance matrix with $\boldsymbol \tau_{\bf \Psi} \bf{Var}(\bf{X})$ where $\bf{Var}(\bf{X})$ denotes the empirical covariance matrix of $\bf{X}$. The matrix $\tau_{\bf \Psi}$ provides the weights of the weighted means, which are used to interpolate the covariance values between sampled and unsampled sites based on the covariance values between sampled sites.

More generally, we can estimate local similarities by multiplying the weight matrix $\boldsymbol \tau_{\bf \Psi}$ with a similarity matrix between sampled sites. In the RESULTS section, we consider similarity matrices that are not correlation or covariance matrices, and we use the pairwise matrix of $(1-F_{ST})$ values for instance. When using individuals as operational units, numerical problems can arise if they are multiple individuals by site because the matrix  $\Psi_{xx}$ can be difficult to invert. Potential solutions are to consider a population---with one or more individuals---at each sampling site or to add a small perturbation to the geographical coordinates of the individuals.

Providing the correlation instead of the covariance between sampled and unsampled sites requires the standardization of the covariance equation (\ref{eq:covariance}) and the renormalization formula is provided in Appendix A.  The final estimate for  the covariance matrix is finally obtained by averaging equation (\ref{eq:covariance}) over posterior replicates of $\boldsymbol \tau_{\bf \Psi}$. The parametric model for ${\bf \Psi}$, which is needed to  generate the posterior distribution of $\boldsymbol \tau _{\bf \Psi}$, is given below.

\subsection*{A model for the correlogram}

To compute the weight matrix $\boldsymbol \tau_{\bf \Psi}$, we consider the standard model of {\it stationary} kriging that assumes that the correlation between two points only depends on the distance between these two points. Using these assumptions, we should model how the correlation decreases with increasing distance. We assume that this function $C$, called the correlogram, decays exponentially
\begin{equation}
\label{eq:decay}
C(d)=((1-\alpha) + \alpha e^{-d/r} + \lambda \mathbbm{1}_{d=0})/(1+\lambda),
\end{equation}
where $d$ is the distance between two points, $\mathbbm{1}$ denotes the indicator function, $\alpha$ determines the sill, which measures the limiting value of the correlation function, $r$ is the {\it range} parameter and $\lambda$ is the {\it regularization} parameter. The parameter $\lambda$ is introduced for numerical reasons because it ensures that the matrix ${\bf \Psi}_{xx}$ is invertible, which is required for the computation of the weight matrix $\boldsymbol \tau_{\bf \Psi}$  \cite[]{bishop06}. The range parameter $r$ is inversely related to the rate at which correlation decays with distance. 
Denoting by $d_{ij}$ the geographical distance between the  $i^{\rm th}$ and $j^{\rm th}$ sites, then the entry of $\bf {\Psi}$ at the $i^{\rm th}$ row and $j^{\rm th}$ column is given by $C(d_{ij})$. We sample the triplet $(\alpha,\lambda,r)$ from the posterior distribution using a MCMC algorithm that contains both Gibbs and Metropolis-Hastings updating steps, and the details of the algorithm are provided in Appendix B \cite[]{Handcock93}. 

\subsection*{Hypothesis-testing procedure}
We introduce two test statistics to test if the variation of local genetic differentiation is significant. The first test statistic is the coefficient of variation of local genetic differentiation values, that is the ratio between standard deviation and mean of local differentiation measures. The second statistic is the distance correlation statistic and it measures the dependence between local genetic differentiation and geographical coordinates. The distance correlation statistic  extends Pearson correlation coefficient because it can measure non-linear dependence \cite[]{szekely07}. Because  we use two test statistics, we consider the conservative Bonferroni correction and reject stationarity when one the two observed values of the test statistics is larger than the $97.5\%$ quantile obtained for the null distribution of stationarity.

We consider two options for generating distribution of the test statistic under the null hypothesis. In the first option, we consider the parametric model of equation (\ref{eq:decay}). We compute $M$ pairwise covariance matrices $\bf{Var}(\bf{X})^i$, $i=1,\dots,M,$ using the stationary correlogram of equation (\ref{eq:decay}). The parameters $(\alpha_i,\lambda_i,r_i)$ of equation (\ref{eq:decay}), which are used to compute the covariance matrices, are sampled according to the posterior distribution. The $2\times M$ values of the tests statistics are then obtained after running the MCMC algorithm  (appendix B) $M$ times for each of the simulated covariance matrix $\bf{Var}(\bf{X})^i$, $i=1,\dots,M$. When the sample size is too large, we have to limit the computational burden of the procedure, and we do not perform $M$ MCMC runs. Instead, for the $i^{\rm th}$ covariance matrix $\bf{Var}(\bf{X})^i$, we use the $i^{\rm th}$ triplet $(\alpha_i,\lambda_i,r_i)$ to compute the weight matrix $\boldsymbol \tau^i_{\bf \Psi}$ and to obtain values of local genetic differentiation. However, equation (\ref{eq:decay}) is only an approximation of the correlation pattern found for isolation-by-distance models. It is exact, for instance, in the one-dimensional stepping-stone model with infinite range \cite[]{kimuraweiss64}. To avoid the approximation of equation (\ref{eq:decay}), we also consider explicit simulations of a stationary stepping-stone model using {\it ms} \cite[]{hudson02}. We consider uniformly sampled migration rates such as $1\leq 4 N_0 m \leq 20$  and we choose a sampling scheme that mimics the sampling of the data.

\section*{Results}
\subsection*{Simulation study}

In the simulation study, we consider two different models for generating non-stationary patterns of isolation-by-distance. First, we consider non-homogeneous stepping stone models in one and two dimensions. We simulate with {\it ms} 2,000 independent SNPs using spatially-dependent effective migration rate $4 N_0 m $ where $N_0$ is the population size of each deme and $m$ is the migration rate per generation between two neighboring demes. Because we assume independence between SNPs, each SNP is simulated with a coalescent simulation that is conditioned on having one segregating site. The second model is analytic and has been developed for performing non-stationary kriging when the correlogram function (equation (\ref{eq:decay})) is assumed to vary across space \cite[]{paciorekschervish06}. The range parameter $r$ of equation (\ref{eq:decay}), which measures the rate at which correlation decays with distance, is assumed to be a function of space.  For the second model, zones of abrupt changes such as genetic barriers correspond to regions with a smaller range parameter because correlation decays more rapidly with distance in these regions.

\subsubsection*{Barrier in a one-dimensional model}
We investigate an example of a one-dimensional model with a genetic barrier. 
We simulate a stepping stone model with 100 populations of effective sizes $N_0=1000$ diploid individuals. Depending on the simulations, we sample either 20 equidistant populations or 20 uniformly sampled populations. We consider 20 chromosomes in each of the population. Migrations are constant between neighboring populations and we consider $4 N_0 m = 4$ and $4 N_0 m = 20$. The barrier is located between populations 50 and 51 and arose 8 units of time ago ($4 N_0 m=0$) where time is counted in units of $4N_0$ generations. As similarity matrix, we consider the pairwise correlation of allele frequencies for  20 sampled populations. For each sampled deme, local genetic differentiation corresponds to one minus the expected correlation between the sampled deme and its two neighbors.

\begin{figure}[ht!]
	\centering
		\includegraphics[width=14cm]{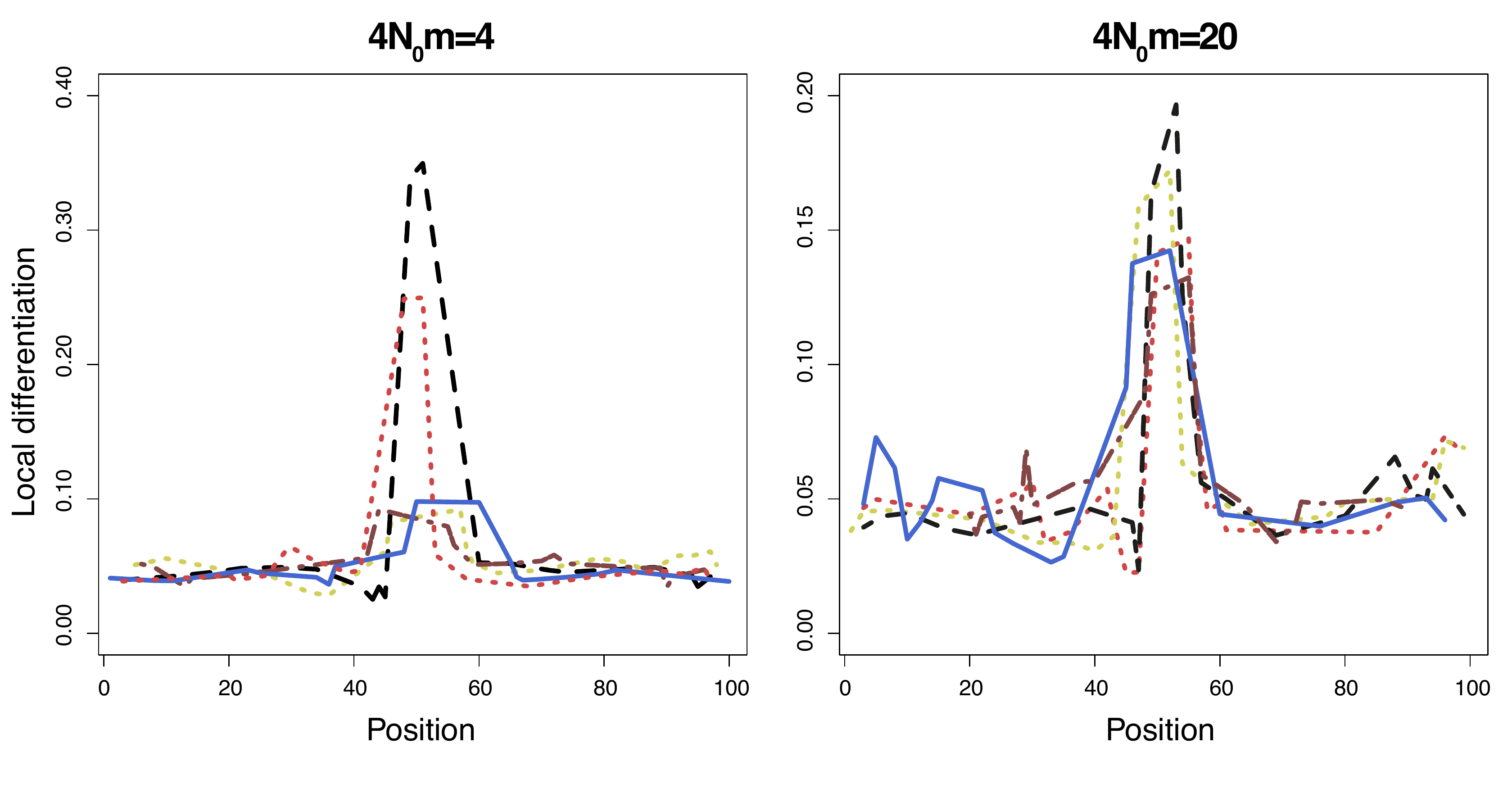}
	\caption{Estimation of local genetic differentiation for five population sets where each set contains 20 randomly-picked populations among 100 populations. Simulations are performed in a one-dimensional stepping-stone model with a barrier to gene flow in the middle of the range ($4 N_0 m=20$ or $4 N_0 m=4$ except between population 50 and population 51 where $4 N_0 m=0$). Local genetic differentiation corresponds to one minus the expected correlation of allele frequencies between sampled demes and their two closest neighboring demes.}
 \label{fig:5sampling} 
\end{figure}

With equidistant sampling, we find that the parameters of the correlogram function (equation (\ref{eq:decay})) affect the estimated values of local genetic differentiation (Figure S1).  However, for all values of the correlogram parameters $(\alpha, \lambda, r)$ we consider, local differentiation is larger in the middle of the range, which is consistent with the presence of a barrier to gene flow. Nonetheless the detailed trajectory of local differentiation as a function of space depends on the correlogram parameters and edge effects can be large for some parameter values (Figure S1). To account for the uncertainty associated with the parameters of the correlogram function, we integrate the values of local genetic differentiation over the posterior distribution of $(\alpha, \lambda, r)$ (Figure S1). 
To investigate if a barrier to gene flow is still detectable with irregular sampling, we also sampled randomly 20 populations among the 100 populations. For both intensities of barrier ($4 N_0 m=20$ or $4 N_0 m=4$ except at the barrier where $4 N_0 m=0$) and for each replicate of population sampling, local differentiation is larger around the barrier to gene flow (Figure \ref{fig:5sampling}). However, for the less stringent and more difficult to detect barrier, local differentiation increases less markedly around the barrier when sampling in the vicinity of the barrier is sparse (Figure \ref{fig:5sampling}).



To provide a comparison, we apply multidimensional scaling (MDS), which is a commonly used method to represent differentiation between populations. Based on the pairwise matrix of correlation between allele frequencies computed for each of the 20 equally-spaced populations, we apply MDS. Figure \ref{fig:mds_1d} displays the scatter plots of the first two principal coordinates when $4 N_0 m = 4$, $4 N_0 m = 20$, and there are 20 equally-spaced sampled populations. In the scenario where $4 N_0 m = 20$, the occurrence of a barrier of gene flow in the middle of the range is visible in the MDS plot whereas it is much less visible when $4 N_0 m = 4$ even with the perfectly regular sampling of populations. In the latter scenario, the MDS plot is above all influenced by the global isolation-by-distance pattern that generates the inverted U-shaped pattern \cite[]{novembreandstephens08}. Patterns exactly similar to MDS were obtained with principal component analysis when using the population allele frequencies as raw data.

We additionally explore the running time of the algorithm. Once a pairwise matrix of $F_{ST}$ or of other dissimilarity measures has been obtained, the most costly operations are the inversion of the matrix ${\bf \Psi}_{xx}$ required to compute the weight matrix $\boldsymbol \tau_{\bf \Psi}$ as well as the computation of a determinant required to evaluate the likelihood in the MCMC algorithm (see Appendix B). The computing cost of both operations is proportional to the cube of the number of sampled sites. We check this theoretical prediction in this example by increasing the total number of demes and the number of sampled demes. We find that the cubic prediction is quite accurate although a bit pessimistic because the running time of the algorithm actually grows as the number of sampled sites at the power 2.5 (Figure S2).

\subsubsection*{Barriers in a two-dimensional model}
\begin{figure}[ht!]
	\centering
		\includegraphics[width=14cm]{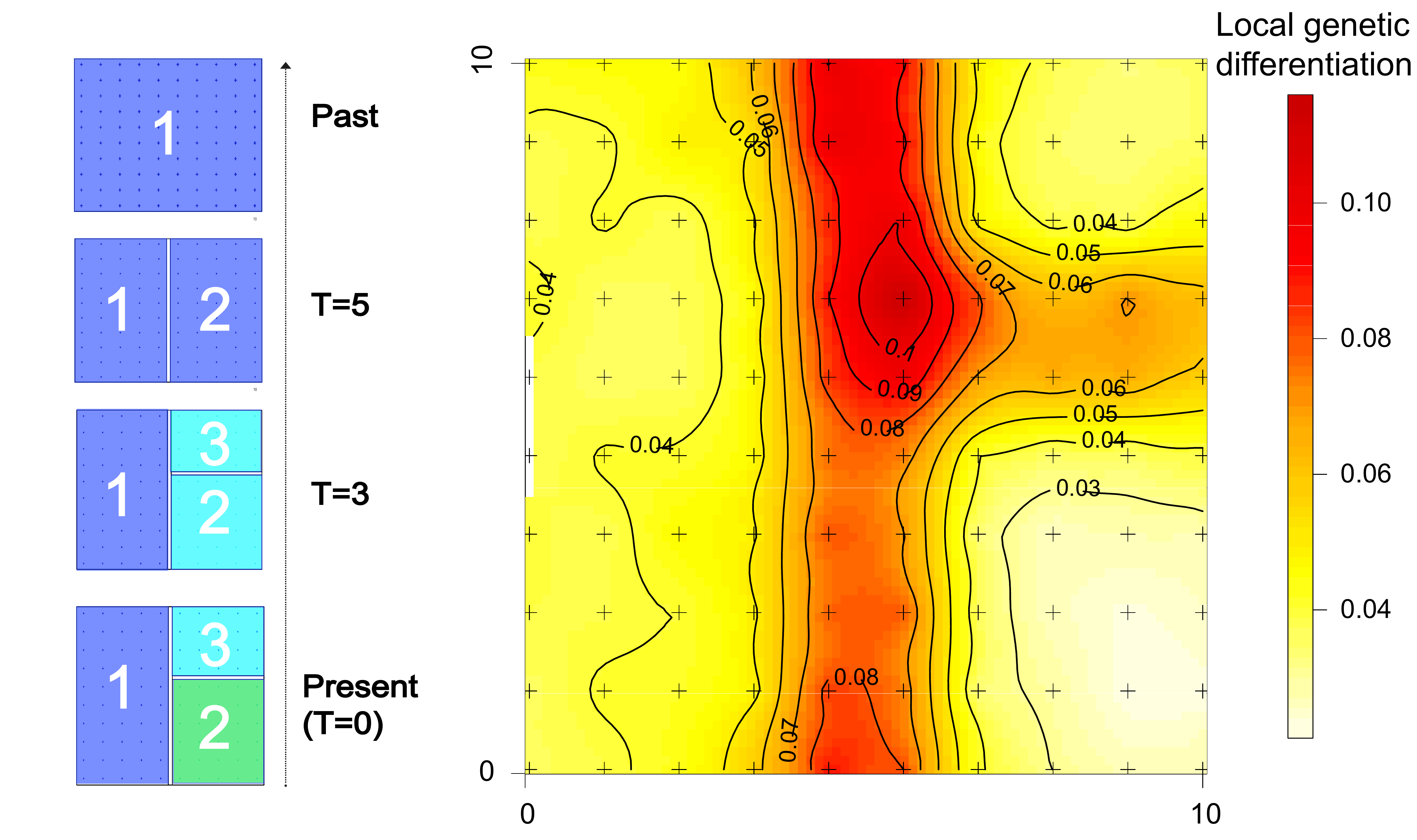}
	\caption{Map of local genetic differentiation in a two-dimensional stepping-stone model with two genetic barriers. On the left-hand side, the time line shows the time when the genetic barriers appeared. Local genetic differentiation was estimated using the pairwise correlation matrix, and it measures one minus the expected correlation between sampled populations and unsampled neighbors located at a distance of 0.1 from the sampled populations.}
\label{fig:2d_ms}  
\end{figure}

We consider two examples of a 2-dimensional model with genetic barriers. In the first 2-dimensional example, the data are simulated using a stepping-stone model in a $10 \times 10$ grid. In each population, there are $N_0=1,000$ diploid individuals per population and we sample all populations considering 20 chromosomes in each of them. The migration rate between neighboring populations is of $4 N_0m = 20$ where migration only  occurs along horizontal and vertical lines but not along diagonals. We then assume that two barriers arose $T_1 = 5$ and $T_2 = 3$ units of time ago where time is counted in units of $4 N_0$ generations (see Figure \ref{fig:2d_ms}). In the second 2-dimensional example, we specify explicitly the local decay of correlation using the non-stationary model of \cite{paciorekschervish06}. We assume that there are three genetic barriers, which correspond to three different regions where the range parameter ($r$ in equation (\ref{eq:decay})) is smaller (Figure S3). In both examples, the input matrix of similarity is the correlation matrix between sampled sites, although the correlation is estimated based on simulated allele frequencies for the stepping stone example whereas it is obtained analytically using the convolution formula of \cite{paciorekschervish06} for the second example.

Figure \ref{fig:2d_ms} and Figure S3 show that estimated values of local genetic differentiation are larger around the genetic barriers as expected. For both examples, the relative importance of the barriers is retrieved. The strongest barrier has the largest value of local genetic differentiation. For computing local genetic differentiation, we additionally consider $F_{ST}$ pairwise values instead of correlation values in the stepping-stone model. In that case, the similarity matrix  contains the pairwise $(1-F_{ST})$ values that decay with increasing geographical distance as assumed by equation (\ref{eq:decay}). The  local genetic differentiation  values now correspond to the expected $F_{ST}$ between the sampled populations and  their fictive neighbors. We find that the map of local genetic differentiation obtained with the $F_{ST}$ measure is similar to the map obtained with the pairwise correlation matrix (Figure S4). We also  perform additional computations of local differentiation after incomplete sampling of the populations in the stepping-stone model. We sample respectively $50\%$, $33\%$, and $25\%$ of the 100 populations present in the grid. We find that the oldest barrier is always recovered but not the mot recent one, which is not detectable when sampling $33\%$ or $25\%$ of the populations (Figure S5).

\subsubsection*{A gradient of gene flow}
\begin{figure}[ht!]
	\centering
		\includegraphics[width=14cm]{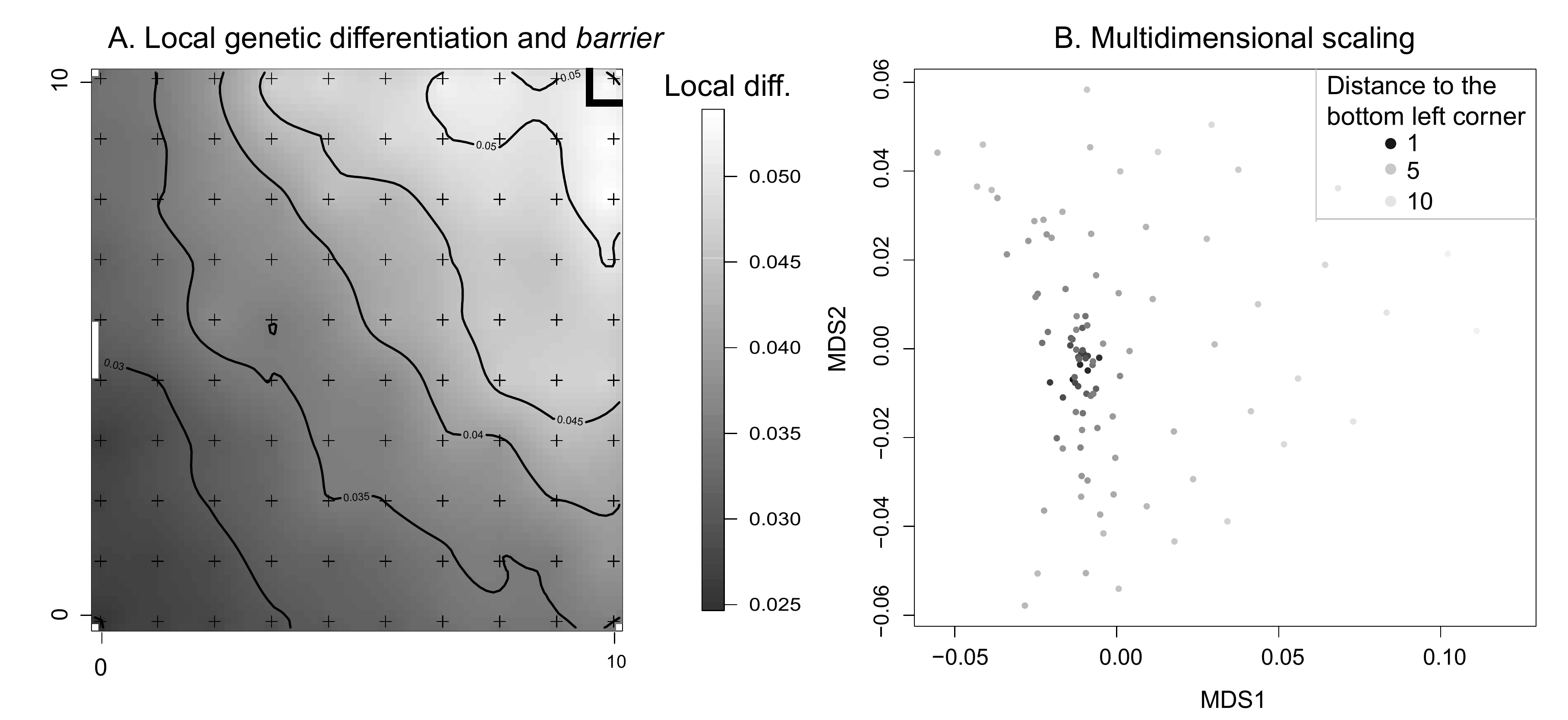}
	\caption{Investigating the pattern of population differentiation for a gradient of gene flow in a two-dimensional model. Gene flow is maximal at the lower left corner of the habitat and decreases proportionally to the distance from the lower left corner. Panel A) Estimation of local genetic differentiation values using the correlation with neighbors located at a distance of $0.1$. The first genetic barrier that is found with the software {\it barrier} is shown with a thick black line. Panel B) Multidimensional scaling plot with a grayscale color scheme to represent the distance of each population to the lower left corner. Dark points correspond to populations that are close to the lower left corner whereas the lightest points are the farthest away from the lower left corner.}
 \label{fig:gradient} 
\end{figure}

We also consider a different pattern of non-stationary IBD consisting of a 2 dimensional stepping-stone model with a spatial gradient of gene flow. We assume that gene flow is maximum at the lower left corner of the habitat and decreases quadratically with distance from the lower left corner of the habitat (Figure \ref{fig:gradient}). As expected, we find that local genetic differentiation increases when moving away from the lower left corner of the habitat (Figure \ref{fig:gradient}). When sampling $50\%$, $33\%$, or $25\%$ of the 100 populations present in the grid, we also find a gradient of local genetic differentiation (Figure S6).

\begin{figure}[ht!]
	\centering
		\includegraphics[width=14cm]{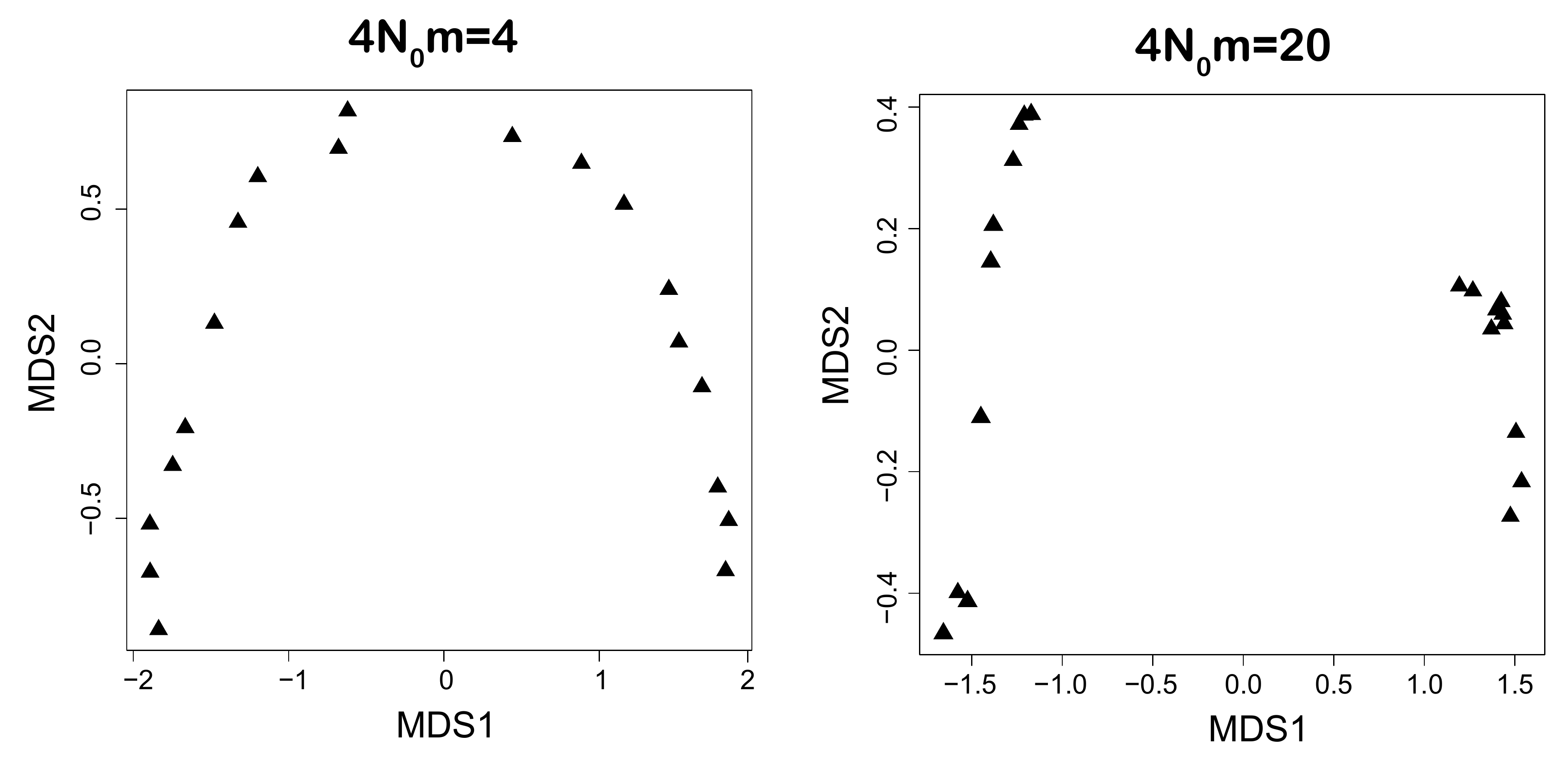}
	\caption{Multidimensional scaling plots  in a one-dimensional range with a genetic barrier in the middle of the range.}
 \label{fig:mds_1d} 
\end{figure}

This example is an instance of non-stationarity, which can not be described with barriers to gene flow. For the previous examples, the software  {\it barrier} \cite[]{mannietal04}, which detects zones of abrupt genetic change, is able to find barriers in both the one and two-dimensional model (Figure S7). However, for the gradient of gene flow,  {\it barrier} incorrectly finds a barrier in the upper right corner of the habitat, which is nonetheless  consistent with the fact that gene flow is minimal here (Figure \ref{fig:gradient}). Additionally, multidimensional scaling provides a meaningful representation for the examples of barriers in the one-dimensional model with $4N_0m=20$ (Figure \ref{fig:mds_1d}) and in the 2-dimensional model (Figure S8) because the populations that live on the same side of the barriers cluster together in the MDS plot (but see $4N_0m=4$ in Figure \ref{fig:mds_1d} where the clustering is less evident). However, interpreting the pattern obtained with MDS is much more difficult for the example of a gradient of gene flow (Figure \ref{fig:gradient}). The observed pattern found with MDS is consistent with the gradient of gene flow because populations living in regions of high gene flow  (dark points in Figure \ref{fig:gradient}) are located more closely on the MDS plot than  populations living in regions of low gene flow (clear points in Figure \ref{fig:gradient}). Although consistent with a gradient of gene flow, the MDS plot is not as easily interpretable as the map of local genetic differentiation for this example.

\subsubsection*{Testing non-stationary patterns of IBD}
\begin{figure}[ht!]
	\centering
		\includegraphics[width=14cm]{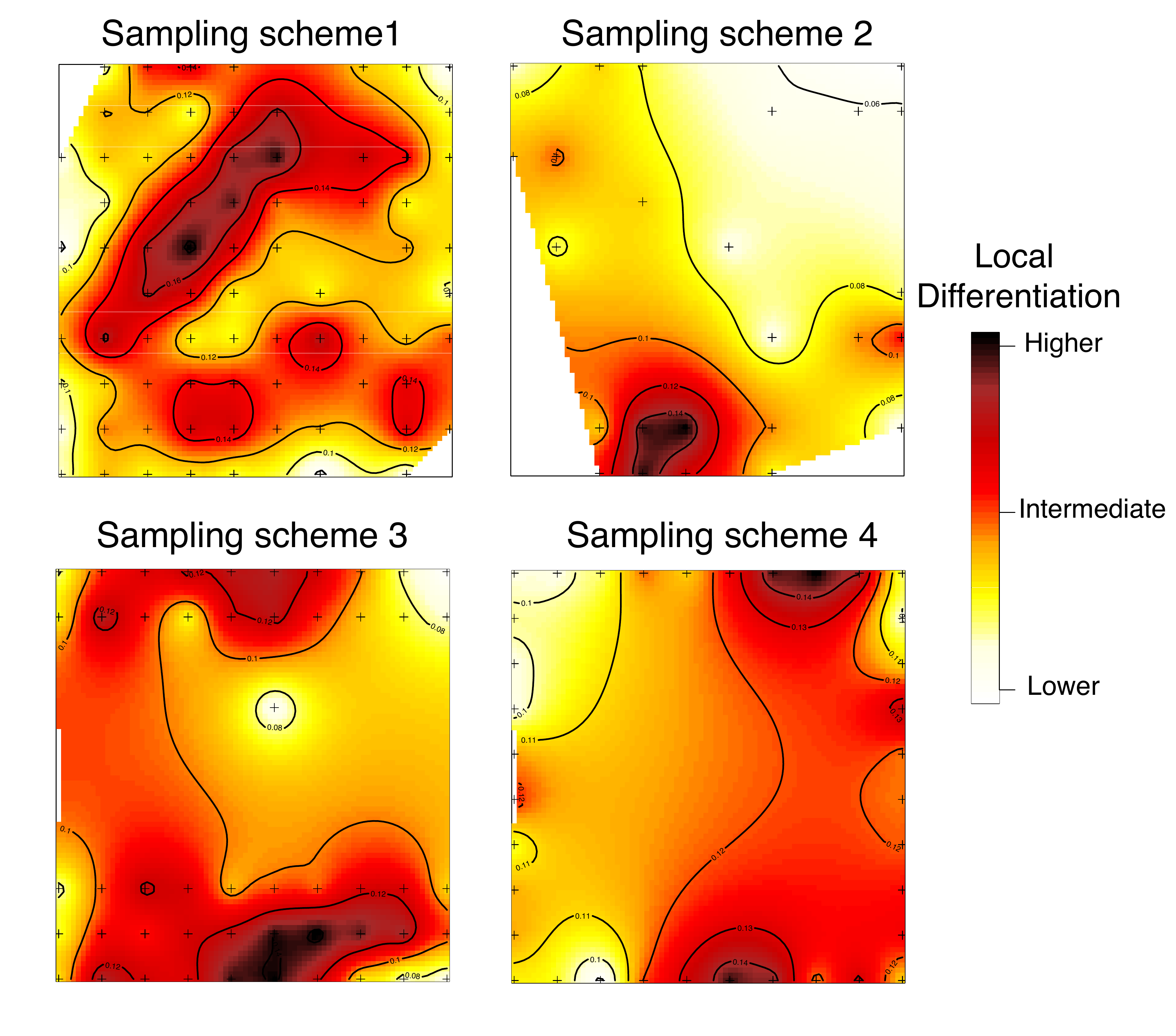}
	\caption{Effect of sampling scheme on local differentiation for data simulated with a stationary model of isolation-by-distance. Simulations are performed with {\it ms} under a stepping-stone model.}
 \label{fig:sampling} 
\end{figure}

The estimates of local genetic differentiation may depend on the sampling scheme. Clustered or irregular sampling scheme in particular can be a matter of concern because they might generate false positive patterns of non-stationarity. Here, we consider different sampling schemes in a 2-dimensional stepping-stone model in order to study the risk of false positives. In the first  and second sampling schemes, respectively $25\%$ and $75\%$  of the total number of sites have been sampled (Figure \ref{fig:sampling}). The third sampling scheme consists of a clustered sampling scheme with two different geographic zones, which have been sampled, as well as an isolated sampled site between the two regions. In the last sampling scheme, only the perimeter of  the two-dimensional square has been sampled. Figure \ref{fig:sampling} shows heat maps of local genetic differentiation for the stationary IBD simulations that have been generated with {\it ms}. To evaluate whether the observed variations of local genetic differentiation are sufficient evidence for non-stationarity, we perform 100 simulations of stationary patterns for each sampling scheme. When simulations of the null models are performed with {\it ms}, stationarity is rejected for $3\%-7\%$ of the simulations, which is consistent with the nominal $5\%$ type I error we use. However if equation (\ref{eq:decay}) is used as a null model, stationarity is rejected for all the simulations performed with {\it ms}. The stepping-stone simulations of stationary IBD show that equation $(\ref{eq:decay})$ should not be used for hypothesis-testing and we should instead resort to explicit simulation of IBD models for generating distributions of the test statistics under the null hypothesis of stationarity. In addition, for all the simulations of non-stationary processes considered so far (barriers in one and two-dimensional models and gradient of gene flow), we reject stationarity as expected.

\subsection*{Applications}
\subsubsection*{Non-stationary patterns of IBD among the Swedish population}
We first illustrate the kriging methodology using a human SNP data set with a particularly dense geographic sampling. The data consist of genome-wide SNPs for 5,174 Swedish individuals that cover all of the 21 Swedish counties \cite[]{humphreysetal11}. To assign each individual to a county, \citet{humphreysetal11} used available geographic information with the following order of priority: city or village of birth, county of birth, municipality or city of residence and county of residence if it is the only information available. They found strong differences between far northern counties and remaining counties, and also showed that northern counties are more clearly genetically differentiated from each other than southern counties are from each other.

\begin{figure}[ht!]
	\centering
		\includegraphics[width=14cm]{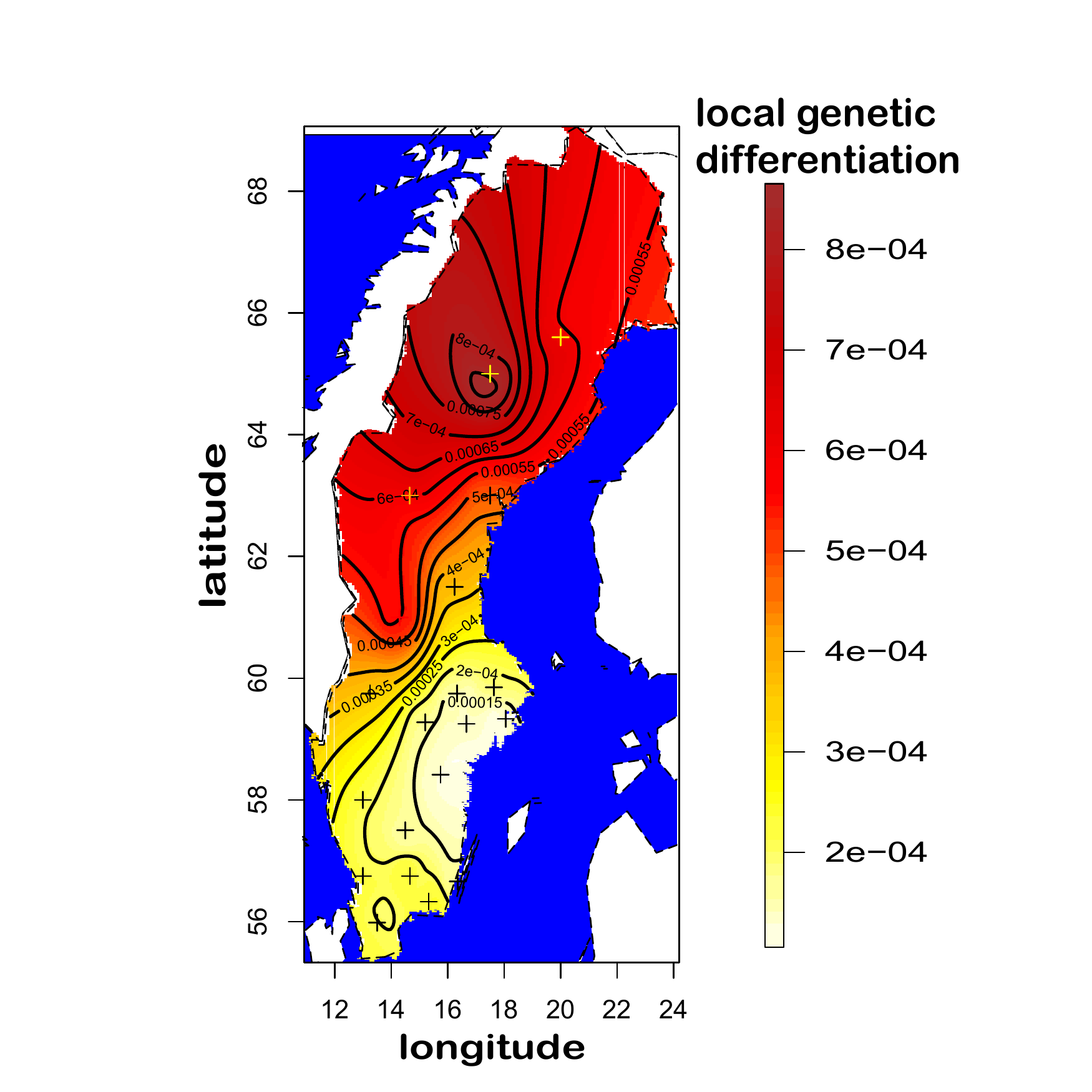}
	\caption{Map of  local genetic differentiation computed for the 20 Swedish counties. Measures of local genetic differentiation correspond to the $F_{st}$ between the sampled counties and fictive neighboring populations, not shown, located at 30 km.}
 \label{fig:sweden} 
\end{figure}

Since our framework is an extension of isolation-by-distance, we first check that population differentiation increases with increasing geographical distance. We confirm the prevalence of  isolation-by-distance in Sweden ($P<10^{-7}$ for a Mantel test, see also Figure S9). Then, we choose to quantify local genetic differentiation using the $F_{ST}$ between a population living exactly in the barycentric center of the county and a putative neighboring population living 30 km away (see Figure S10). We consider the pairwise $(1-F_{ST})$ values between the counties as input matrix of pairwise similarities. Our hypothesis-testing procedure indicates significant non-stationary pattern of IBD in Sweden. We find that the northernmost counties (Nordbotten, V\"asterbotten and J\"amtlands) have the strongest values of local genetic differentiation (Figure \ref{fig:sweden}) whereas the smallest values are found in the regions around the Stockholm area (\"Osterg\"otlands, Stockholms, S\"odermanlands, V\"astmanlands and J\"onk\"opings are the five counties with the lowest  values of local differentiation). The fourth largest value is found for the Dalarna county (in north middle Sweden), which borders southern Norway, and counts more individuals with remote Finnish or Norwegian ancestry than other counties \cite[]{humphreysetal11}. As expected, the four counties with the largest local differentiation values are also the most differentiated from other counties even when controlling for geographic distance (Figure S9).

In summary, we confirm the results of \cite{humphreysetal11} who found that there is more genetic differentiation within northern Sweden than within southern Sweden. The four counties with the largest values of local genetic differentiation are also the counties with the lowest population densities (Figure S11) suggesting that low population density triggered population differentiation in northern Sweden.

\subsubsection*{Non-stationary patterns of IBD for alpine plant species}
We consider a set of 20 alpine plant species that have been sampled across the Alps \cite[]{gugerlietal08,alvarezetal09,jayetal12}. The sampling is particularly dense with one to three individuals per species collected for each cell of approximately 500 ${\rm km}^2$. Individual genotypes consist of AFLPs. We compute allele frequencies at each sampling site--possibly using one individual only--and we consider the matrix of correlation between allele frequencies as input similarity matrix. Local genetic differentiation corresponds to one minus the expected correlation between sampled populations and neighboring populations located at $8$ km.

\begin{figure}[ht!]
	\centering
		\includegraphics[width=18cm]{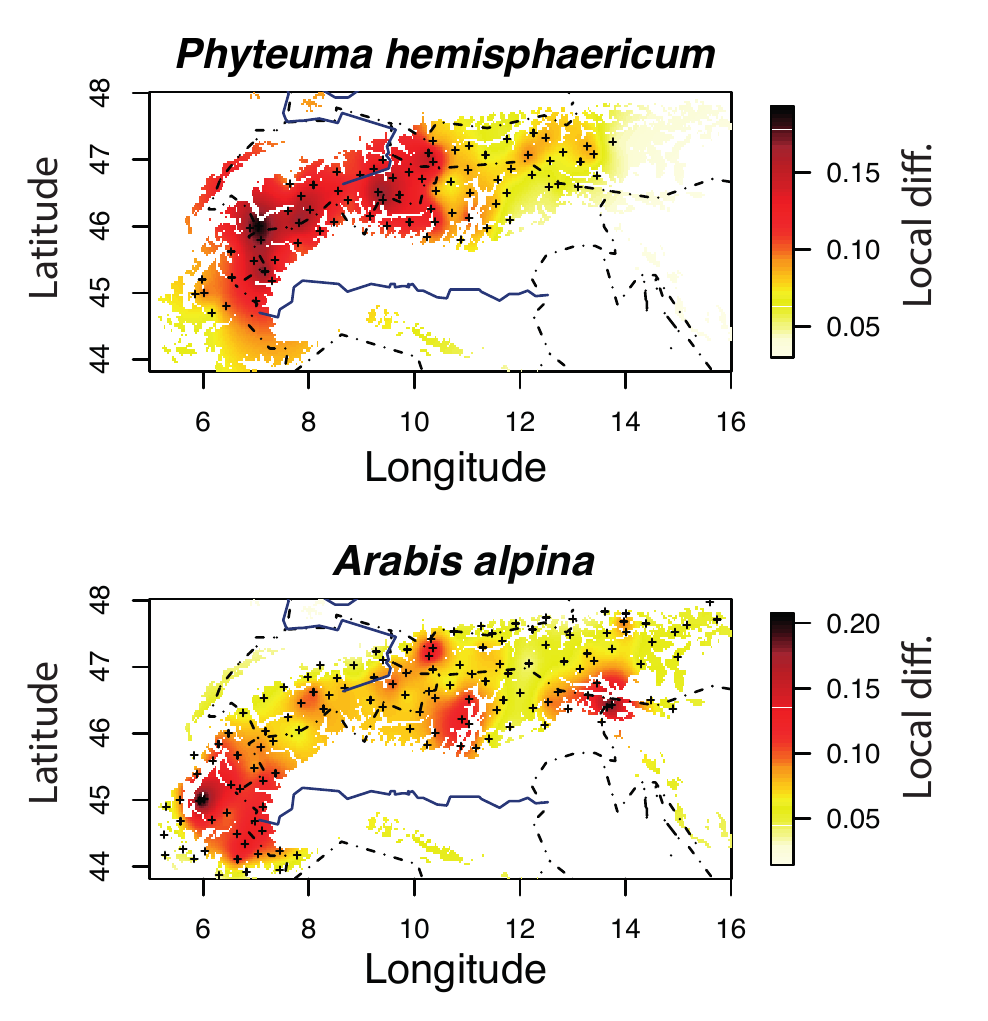}
	\caption{Heatmap of local genetic differentiation computed across the Alps for two alpine species: {\it Phyteuma hemisphaericum} and {\it Arabis alpina}. Local differentiation corresponds to one minus the correlation between sampled populations and fictive neighboring populations, not shown, located at 8 km around the sampling sites.}
 \label{fig:alpine} 
\end{figure}

The test  for non-stationarity is significant for 7 species with a type I error rate of $5\%$ and it increases to 9 species when accepting a type I error rate of $10\%$ (Table S1). However, for the species with non-stationary IBD,  the detailed pattern of local genetic differentiation is idiosyncratic to each species (Figure S12). For instance, the alpine species {\it Phyteuma hemisphaericum}
exhibits larger values of local genetic differentiation in a large region ranging from the central Alps to the southwestern Alps, whereas  there are disconnected regions of larger genetic differentiation for the species {\it Arabis Alpina} all located in the South of the Alps (Figure \ref{fig:alpine}). To integrate the results found for all species with non-stationary IBD, we compute, for each species, normalized rank values between 0 and 1 where 0 corresponds to the site of lowest local differentiation and 1 corresponds to the site of largest local differentiation. When averaging the normalized ranks across species, we found that a region of the Western Alps  encompassing the  inner alpine Aosta valley is the region with the larger values of local differentiation (Figure S13). 
This region has already been found to be one of the two major break zones of allele distribution patterns for alpine plant species \cite[]{thieletal2011}. Pleistocene glaciations are putative explanations for the occurrence of a break zone in this region: the populations of plants were initially fragmented into glacial refuges, then expanded via postglacial colonization routes, and a secondary contact zone finally arose where formerly allopatric populations admixed \cite[]{pawlowski70,schonswetteretal05,thieletal2011}.

\section*{Software}

The software {\it LocalDiff} implementing the method is available at  \url{http://membres-timc.imag.fr/Michael.Blum/LocalDiff.html}. It computes local genetic differentiation from a matrix of pairwise similarity score and can also handle raw genotype data that contain genotypes of individuals. In addition, it generates the {\it ms} command lines that are required for performing the stepping-stone simulations used in hypothesis-testing.

\section*{Discussion}

In this article, we present a new Bayesian method to characterize non-stationary patterns of isolation-by-distance. Because the method aims at refining the description of isolation-by-distance patterns, it should only be applied when isolation-by-distance has already been detected. From {\it global} measures of pairwise similarity or dissimilarity, the method infers {\it local} measures of similarity or dissimilarity. Whatever is the exact measure of genetic (dis)similarity, we use the generic expression of {\it local genetic differentiation} when referring to the estimated  local growth of genetic dissimilarity or differentiation. If considering  for instance the $F_{ST}$ pairwise matrix of  genetic differentiation between populations, the inferred values correspond to the $F_{ST}$ between the sampled populations and fictive neighboring populations located at a given distance. The method is not restricted to $F_{ST}$ measures and can handle any type of measures of differentiation and is also valid at the individual scale. We consider for instance the correlation between the allelic types of individuals, but other measures would be valid such as identity by descent between individuals \cite[]{browningbrowning11} as well as coancestry measures \cite[]{lawsonetal12}. Because the two latter measures are based on haplotypes instead of genotypes, they can provide information at a finer geographical scale \cite[]{gattepaillejakobsson12,lawsonetal12}. 

\subsection*{Genetic differentiation and gene flow}
It is of course tempting to convert maps of  local genetic differentiation into maps of gene flow or of dispersal distance. Assuming that differentiation occurs according to a stepping stone model, such parameter estimates could be obtained using theoretical relationships between local $F_{ST}$ and dispersal distance \cite[]{rousset97}. However, relating $F_{ST}$ or other measures of genetic differentiation to gene flow relies on many assumptions that may be unrealistic \cite[]{markoandhart11}. Although the estimation of gene flow with $F_{st}$-based methods can be robust in some situations, such as temporal variation of gene flow \cite[]{lebloisetal04}, there are other processes such as range expansion, local extinction and recolonization that can modify drastically the pattern of genetic differentiation \cite[]{wadeandmccauley88,arenasetal12}. More generally, a map of local genetic differentiation is informative about the {\it pattern} of genetic differentiation but do not provide enough information to distinguish between the possible evolutionary {\it processes} that generated this pattern \cite[for similar concerns about PCA, see][]{mcvean09}. To provide a concrete example, the same pattern, a zone of elevated local differentiation, can be interpreted as a barrier to gene flow in a equilibrium stepping-stone process (Figures \ref{fig:5sampling}-\ref{fig:2d_ms}) or as a secondary contact zone following postglacial expansions in the case of the alpine plant species  \cite[]{thieletal2011}. 

\subsection*{Relevance to landscape genetics}
Two key steps in landscape genetics are the detection of genetic discontinuities and the correlation of these discontinuities with landscape and environmental features such as barriers \cite[]{maneletal03}. Detection of genetic discontinuities is clearly provided by the proposed kriging method; for instance in the case of the alpine species {\it Phyteuma hemisphaericum}, we find genetic discontinuities, i.e. larger local genetic differentiation, in a large region of the central and western Alps (Figure \ref{fig:alpine}). However, aiming at capture  genetic discontinuities using barriers only might be too limited and the kriging method can reveal more complex patterns such as gradient of local differentiation across the species' range (Figure \ref{fig:gradient}). The second key step where the genetic discontinuities are correlated with landscape or environmental variable can also be obtained as a post-processing step by correlating estimated local genetic differentiation with landscape variables. For instance, in the case of the human SNP  Swedish data, we find that local genetic differentiation is correlated with population density. There are alternative and {\it integrative} approaches that account for both genetic data and landscape variables within the same statistical framework. Accounting for both sources of data can be performed either by a joint assessment of the pattern of population structure or differentiation and its correlation with landscape or environmental variable \cite[]{follandgaggiotti06,jayetal11}, or by correlating genetic distances with distances based on landscape features \cite[]{cushmanetal06,mcrae06}. These integrative approaches are {\it hypothesis-driven} in the sense that each set of landscape features affecting population structure corresponds to one hypothesis that can be tested or compared to other ones. The proposed kriging approach is instead a technique of {\it exploratory data analysis}. It might be especially appropriate for large-scale conservation studies not focused on the underlying evolutionary processes but that should deal with reserve design and with the management of fragmented populations \cite[]{schwartzetal2007}. To explore patterns of genetic differentiation, there are other statistical summaries of the data that can be computed. Population-specific $F_{ST}$'s based on the $F$-model can also provide local measures of differentiation by computing local values of $F_{ST}$'s \cite[]{GaggiottiFoll10}. However, compared to approaches based on the $F$-model, {\it LocalDiff} can also work with the individual-based sampling schemes often encountered in landscape genetics \cite[]{schwartzmckelvey09}. Moreover, $F_{ST}$'s based on the $F$-model relies on a parametric population-genetic model that may be sensitive to departures from the assumption of the $F$-model \cite[]{GaggiottiFoll10}. By contrast, the proposed approach relies on kriging (aka Gaussian process), which is a non-parametric approach that assumes a pattern of isolation-by-distance only.

To compare or test the support of different evolutionary processes provided by the pattern of non-stationary IBD, we can rather resort to inference based on explicit simulations of evolutionary processes using for instance approximate Bayesian computation \cite[]{csilleryetal10}. Within this simulation framework, measures of local genetic differentiation can be included as statistical summaries of the data. When the number of sampling sites is large, the MCMC algorithm might be too slow to provide summary statistics for ABC inference. To overcome this problem, we provide an option in {\it LocalDiff} where measures of local differentiation are computed by integrating the parameter triplet $(\alpha,\lambda,r)$ over the prior distribution instead of the posterior distribution. Although integration over the posterior with MCMC should be preferred when possible, integrating over the prior distribution provide relevant measures of local differentiation for the examples we investigated (see Figure S14).

\subsection*{Caveats}
Although, computing local differentiation is a descriptive technique of exploratory data analysis, we provide an hypothesis-testing procedure attached to it. This is desirable feature since it can prevent from overinterpreting maps of local differentiation (Figure \ref{fig:sampling}). The test of non-stationarity relies on two different test statistics and the null distribution is obtained from stepping-stone simulations performed with {\it ms}. To obtain the null distribution of the test statistic, we consider the same sampling as in the data, which is crucial since the sampling scheme can affect the inferred pattern of local differentiation (Figure \ref{fig:sampling}). For the simulations of the null model, we uniformly sample the parameter $4 N_0 m$ in a fixed range but we acknowledge that using an estimated value of the effective migration parameter would also be a valid strategy. We also consider an alternative and approximate null model under which  the correlation decays exponentially (equation (\ref{eq:decay})) but this approximation should not be used for hypothesis testing because it is much too liberal. However, even considering explicit stepping-stone simulations might have drawbacks because other processes such as range expansions can also generate IBD patterns but they might induce different distributions of tests statistics \cite[]{edmondsetal04}. More generally, sampling scheme affects the different methods that characterize population differentiation and simulations can be informative about the effects of sampling scheme \cite[]{mcvean09,schwartzmckelvey09,jayetal12b}.

Another concern about the kriging method concerns the choice of the distance between the sampled populations or individuals and the fictive neighboring populations or individuals.  We consider various choices of distances for the Swedish data set (10-150 km) and find that all these choices of distance provide similar patterns of local genetic differentiation although the patterns generated with the larger distances are smoother (Figure S15). The last caveat to bear in mind concerns the choice of the pairwise dissimilarity or differentiation matrix. For instance, in a simulation study with 5 populations, \cite{lawsonfalush12} showed that the five populations were clearly distinguishable with some but not all pairwise dissimilarity matrices between individuals. Although a potential caveat, being able to choose the measure of dissimilarity also adds to the flexibility of the method and different measures can convey information about processes that occurred at different time periods.

\subsection*{Perspectives}
When carefully addressing the aforementioned caveats, measures of local genetic differentiation can provide interpretable patterns for describing non-stationary patterns of IBD. Here, the expression {\it non-stationary} refers to spatial variations but note that if temporal data are available, a similar kriging framework can be used to study how genetic drift evolves as a function of time. The present method relies on a matrix of pairwise (dis)similarity between individuals or populations located on georeferenced sampling sites. An approach based on dissimilarity matrices is an appropriate methodology for the new genomic era where we have to deal with massive data. Computing pairwise dissimilarity matrix can be computationally efficient \cite[e.g.][]{browningbrowning11} and can even be parallelized to compute different parts of the matrix \cite[]{lawsonfalush12}. Having a statistical method able to scale with the dimension of the genetic data should make it a valuable tool for investigating patterns of genetic differentiation in a wide range of studies.

\bibliographystyle{evolution}
\renewcommand\refname{Literature Cited}
\bibliography{friction}

\clearpage

\appendix
\section*{Appendix A: Normalizing the covariance matrix} \label{sec:cor}
To renormalize equation (\ref{eq:covariance}) and compute the correlation matrix, we have to compute the diagonal elements of the variance-covariance matrix $\bf{Var}(\bf{Y})$ for the unsampled sites. Using the fact that the residuals of the regression equation (\ref{eq:reg_krig}) are of variance ${\bf \Psi}_{yy}-{\bf \Psi}_{xy}{\bf \Psi}_{xx}^{-1}{\bf \Psi}_{xy}$ \cite[]{bishop06}, we can show that the variance-covariance matrix $\bf{Var}(\bf{Y})$ is given by
$$
\bf{Var}(\bf{Y})=\tau_{\bf \Psi} \bf{Var}(\bf{X}) \tau_{\bf \Psi}^T + \Psi_{yy}-\Psi_{xy}\Psi_{xx}^{-1}\Psi_{xy}.
$$

\section*{Appendix B:  Gibbs sampler} \label{sec:Gibbs}

%

For the parameters of the correlogram model, we choose the following prior distributions

\begin{equation}
\label{eq:prioralpha}
\alpha \sim \mathcal{U}(\min  ({\bf Var}({\bf X}) )- \delta, \min  ({\bf Var}({\bf X}))+ \delta) 
\end{equation}
\begin{equation}
\label{eq:priorlambda}
\log_{10} \lambda \sim \mathcal{U}(-4, -1)
\end{equation}
\begin{equation}
\label{eq:priorr}
r \sim \mathcal{U}(\min D_{ij}, \max D_{ij}),
\end{equation}
where $\min (\bf{Var}(\bf{X})) $ is the smallest element of the variance-covariance matrix of $\bf{X}$ and $D$ denotes the pairwise geographical distance between sampled sites.
Here, we adopt an empirical Bayes approach, since we partly use the data when defining the prior distributions.

Using the Bayes formula we have
\begin{equation}
\label{eq:bayes}
\log(p(\alpha, \lambda, r | {\bf Var}({\bf X}) )) = \log(p({\bf Var}({\bf X}) | \alpha, \lambda, r)) + \log(p(\alpha, \lambda, r)) + C_1,
\end{equation}
where $C_1$ is a constant and the likelihood is given by the Wishart distribution \cite[]{schmidtandohagan03} 
\begin{equation}
\label{eq:likelihood}
\log(p({\bf Var}({\bf X})| \alpha, \lambda, r)) = -\frac{l-1}{2}\log(|{\bf \Psi_{xx}}|) - \frac{l}{2}{\rm trace}({\bf \Psi_{xx}}^{-1}{\bf Var}({\bf X})) + C_2,
\end{equation}
where $C_2$ is another constant, and $l$ is the number of loci. We then simulate a sample of the joint posterior probability using  an hybrid algorithm with Gibbs and Metropolis-Hastings updating steps. To obtain replicates from the conditional distributions $p(\alpha | \lambda, r , {\bf Var}({\bf X}))$ and $p( \lambda | r, \alpha , {\bf Var}({\bf X}))$ we evaluate the likelihood function over a grid for the triplet $(\alpha, \lambda, r)$. The evaluations of the likelihood function are performed before running the MCMC algorithm and are saved to be subsequently used during the course of the algorithm. We evaluate the conditional densities for each point of the grid using equations (\ref{eq:bayes}) and (\ref{eq:likelihood}) and then we use a sampling algorithm for simulating discrete random variables with known probability masses. The updates for the parameter $r$ are performed with a Metropolis-Hastings (MH) algorithm using as proposal a discrete random walk over the grid. The resolution of the grid is more acute for the parameter $r$ than for the other 2 parameters, which explains why we consider a MH step that was less time-consuming than evaluating the likelihood function for all the points of the grid.

\clearpage


\clearpage


\clearpage

\clearpage

\clearpage

\clearpage


\clearpage

\clearpage


\end{document}